\documentclass[aps, pra, superscriptaddress, floatfix, reprint]{revtex4-1}
\usepackage{graphicx}
\usepackage{dcolumn}
\usepackage{bm}
\usepackage{amsfonts}
\usepackage{amsmath}
\usepackage{amssymb}
\usepackage{color}
\usepackage[normalem]{ulem} 
\usepackage[percent]{overpic}

\usepackage[colorlinks=true,citecolor=blue]{hyperref}
\hypersetup{colorlinks=true,citecolor=blue,linkcolor=red,urlcolor=blue}

\renewcommand{\thefigure}{\arabic{figure}}
\def\be{\begin{equation}}
\def\ee{\end{equation}}

\def\kv{{\bf k}}
\def\qv{{ \bf q}}

\begin{document}
	\renewcommand{\thefigure}{\arabic{figure}}

	\title{Composite quasiparticles in strongly-correlated dipolar Fermi liquids}
	\author{Iran Seydi}
	\affiliation{Department of Physics, Institute for Advanced Studies in Basic Sciences (IASBS), Zanjan 45137-66731, Iran}
	\author{Saeed H. Abedinpour}
	\email{abedinpour@iasbs.ac.ir}
	\affiliation{Department of Physics, Institute for Advanced Studies in Basic Sciences (IASBS), Zanjan 45137-66731, Iran}
	\affiliation{Research Center for Basic Sciences \& Modern Technologies (RBST),  Institute for Advanced Studies in Basic Sciences (IASBS), Zanjan 45137-66731, Iran}
	\affiliation{School of Nano Science, Institute for Research in Fundamental Sciences (IPM), Tehran 19395-5531, Iran}
	\author{Reza Asgari}
	\affiliation{School of Physics, Institute for Research in Fundamental Sciences (IPM), Tehran 19395-5531, Iran}
	\affiliation{School of Nano Science, Institute for Research in Fundamental Sciences (IPM), Tehran 19395-5531, Iran}
	\author{B. Tanatar}
	\affiliation{Department of Physics, Bilkent University, Bilkent, 06800 Ankara, Turkey}
	\date{\today}
	
	\begin{abstract}
		Strong particle-plasmon interaction in electronic systems can lead to composite \emph{hole-plasmon} excitations.
		We investigate the emergence of similar composite quasiparticles in ultracold dipolar Fermi liquids originating from the long-range dipole-dipole interaction.
		We use the $G_0W$ technique with an effective interaction obtained from the static structure factor to calculate the quasiparticle properties and single-particle spectral function.
		We first demonstrate that within this formalism a very good agreement with the quantum Monte Carlo results could be achieved over a wide range of coupling strengths for the renormalization constant and effective mass.
		The composite \textit{quasiparticle-zero sound} excitations which are undamped at long wavelengths emerge at intermediate and strong couplings in the spectral function and should be detectable through the radio frequency spectroscopy of nonreactive polar molecules at high densities.
	\end{abstract}
	
	\maketitle

	\section{Introduction}
	Ultracold dipolar gases thanks to their long-range and anisotropic dipole-dipole interactions are excellent candidates for exploring quantum many-body behavior~\cite{Baranov1, Baranov4, Bloch}. 
	The ground state properties of the two dimensional (2D) dipolar Fermi liquids (DFL) have been widely explored by means of different techniques, namely the quantum Monte Carlo (QMC) simulation~\cite{Matveeva}, modified Singwi-Tosi-Land-Sj\"{o}lander (STLS) method~\cite{Parish}, and Fermi-Hypernetted-Chain (FHNC) approximation~\cite{Abedinpour}. The instability of a homogenous 2D liquid towards density modified phases have been investigated using different formalisms~\cite{Yamaguchi, Zinner, Sieberer}.
	The Landau-Fermi liquid properties of a 2D DFL have been addressed by Lu and Shlyapnikov~\cite{Lu1} and it has been shown that its collective density excitation has an acoustic nature whose survival at long wavelengths originates mainly from the many-body effects beyond the mean-field level~\cite{Lu1, Abedinpour, Lu2013}.
	Invaluable information regarding the excitation spectrum of a correlated many-body system could be obtained from the self-energy and in particular from the single-particle spectral function.
	Dynamics of quantum fluids, and mainly electron liquids have been studied extensively over the past decades.

	Quasiparticle (QP) spectral properties of a 2D electron liquid~\cite{Jalabert} and a single layer of doped graphene~\cite{Hwang, Polini} have been investigated within the so-called $G_0W$ approximation and the composite \textit{hole-plasmon} excitations have been predicted for both systems. These composite quasiparticles are named plasmarons and have been experimentally verified in doped graphene via angle-resolved photoemission spectroscopy (ARPES) measurements~\cite{Bostwick}.
	In addition, inelastic neutron scattering measurements of a monolayer of liquid $^3$He has been reported to observe a roton-like excitation as an unexpected collective behavior of a Fermi many-body system~\cite{Godfrin}.
	
	Although the dynamical properties of 2D dipolar Bose liquids have been theoretically explored~\cite{mazzanti_prl2009, hufnagl_prl2011, macia_prl2012}, to the best of our knowledge such studies for fermionic systems are restricted to perturbative approaches at weak couplings~\cite{Lu2013}. In this work, we use the $G_0W$ approximation with an effective interaction obtained from the interacting static structure factor to calculate the self-energy. We use an accurate interacting static structure factor data  extracted from FHNC approximation~\cite{Abedinpour} to obtain the effective particle-particle interaction. Below, we will first illustrate how an excellent agreement with the QMC data for effective mass and renormalization constant could be achieved with such a formulation of the $G_0W$ approximation.
	The quasiparticle properties and in particular the effective mass obtained from the $G_0W$ is usually very sensitive to the approximations employed for the effective interaction~\cite{Asgari1}. The level of agreement with QMC data we have obtained with this modified $G_0W$ formulation provides a simple but accurate recipe for the investigation of other strongly interacting Fermi liquids.
	Then, we move to investigate the single-particle spectral function of a 2D DFL. Alongside the usual quasiparticle excitation dispersion below the Fermi energy, a secondary heavy mode at intermediate and strong couplings emerges originating from the coupling between QP and zero-sound excitations. The high-density limit necessary for the observation of this composite mode requires nonreactive fermionic polar molecules and radio frequency (RF) spectroscopy~\cite{torma_scripta2016, Gupta2003} could be employed to probe these features.

	\section{Theory}\label{sect:theo}
	We consider a single layer of spin-polarized (i.e., single component) 2D gas of dipolar fermions with their dipole moments aligned in the perpendicular direction to the 2D plane at zero-temperature.
	The isotropic dipole-dipole interaction between particles is $v (r)=D/r^{3}$, where $D$ is the dipole-dipole interaction strength~\cite{Abedinpour}. At zero temperature all properties of this dipolar system will depend on a single dimensionless coupling constant
	$\lambda=k_{\rm F}r_{0}$,
	where $k_{\rm F}=\sqrt{4{\pi}n}$ is the Fermi wave vector at density $n$ and $r_{0}=m D/\hbar^{2}$ is the characteristic length of dipolar interaction, $m$ being the bare (i.e., noninteracting) mass of dipoles.
	We assume low-lying excitations and resort to the $G_0W$ approximation~\cite{Vignale} to calculate the self-energy, $\Sigma (k, E)$,
	\begin{equation}\label{G0_W}
	\Sigma(k,E)=i \int\frac{\mathrm{d}^2{\bf q}\,\mathrm{d}(\hbar\omega)}{(2\pi)^3}
	G^{0}(\textbf{k}-\textbf{q},E-\hbar \omega) W(q,\omega).
	\end{equation} 
	Here, $G^{0}(\textbf{k},E)=1/[E - \varepsilon_0(k)]$ is the noninteracting Green's function with the noninteracting dispersion of single particle given by $\varepsilon_0(k)=\hbar^2k^2/(2 m)$. 
	In order to account for the effects of exchange and correlations, we have replaced $W(q,\omega)$ by the Kukkonen-Overhauser (KO) effective interaction~\cite{Vignale}
	\begin{equation}\label{K-O int}
	W_{\rm KO}(q,\omega) = v(q) + w^2(q) \chi(q,\omega),
	\end{equation}  
	where $v(q)$ is the Fourier transform of bare interaction and $w(q)$ is effective particle-particle interaction and accounts for the exchange and correlation effects~\cite{Vignale}. The $w(q)$ is indeed the screened interaction usually defined in terms of the many-body local field factors, but here, using the fluctuation-dissipation theorem, we obtain an approximate expression for it in terms of the interacting static structure factor $S(q)$, as~\cite{Abedinpour}
	\begin{equation}\label{eq:FHNC}
	w(q)=\frac{\varepsilon_0(q)}{2n}\left[{S^{-2}(q)}-{S_{0}^{-2}(q)}\right].
	\end{equation}
	Here, $S_0(q)$ is the static structure factor of an ideal 2D Fermi system~\cite{Vignale} and for the interacting structure factor we have used the accurate numerical data from FHNC method reported in Ref.\,\cite{Abedinpour}.
	The interacting linear density-density response function $\chi (q,\omega)$ is written in terms of the screened interaction $w(q)$ in a generalized random-phase approximation (RPA) form~\cite{Vignale}
	$\chi (q,\omega)=\chi_{0} (q,\omega)/\left[1-w(q)\chi_{0} (q,\omega)\right]$, where $\chi_{0} (q,\omega)$ is the noninteracting linear density-density response function of a 2D Fermi system~\cite{Vignale}.
	The self-energy is conveniently split into two terms, namely the ``Hartree-Fock" (HF) term $\Sigma_{\rm HF} (k)$, and the remaining dynamic term $\Sigma_{\rho} (k, E)$, which originates from the density-fluctuations.
	The HF contribution could be written as
	\be
	\Sigma_{\rm HF} (k) =  \int\frac{\mathrm{d}^2{\bf q}}{(2\pi)^2}  \left[v(0) - v({\bf k}-{\bf q})\right]n_{\rm FD}[\varepsilon_0(q)],
	\ee
	where $n_{\rm FD}(\varepsilon)$ is the noninteracting Fermi-Dirac distribution function.
	The dynamical contribution to the self-energy
	\be\label{dynamical_term}
	\Sigma_{\rho} (k,E) =  i \int\frac{\mathrm{d}^2{\bf q}\,\mathrm{d}(\hbar\omega)}{(2\pi)^3} G^{0}(\kv-\qv,E-\hbar \omega) w^2(q) \chi (q,\omega),
	\ee
	itself, for numerical conveniences, is usually further split into two contributions,
	a smooth line-term
	\be\label{line-term}
	\begin{split}
		\Sigma_{line}^{(\rho)}(k,E)=
		- \int &\frac{\mathrm{d}^2{\bf q}\,\mathrm{d}(\hbar\omega)}{(2\pi)^3} w^{2}(q) \chi (q,i\omega)\\
		&\times \frac{E-\xi_0(\textbf{k}-\textbf{q})}{[E-\xi_0(\textbf{k}-\textbf{q})]^2 +(\hbar \omega)^{2}},
	\end{split}
	\ee
	with an integral along the imaginary frequency axis, and a pole-term 
	\be\label{sigma_pole}
	\begin{split}
		\Sigma_{pole}^{(\rho)} (k,E) = \int &\frac{\mathrm{d}^2{\bf q}}{(2\pi)^2} w^{2}(q) \chi(q,E-\xi_0(\textbf{k}-\textbf{q})) \\
		&\times\left[\Theta(E-\xi_0(\textbf{k}-\textbf{q}))-\Theta(-\xi_0(\textbf{k}-\textbf{q}))\right],
	\end{split}
	\ee
	originating from the residue of the single-particle Green's function. 
	Here, $\xi_0(k) = \varepsilon_0(k) - \varepsilon_{\rm F} $, where $\varepsilon_{\rm F} = \hbar ^2  k_{\rm F}^2/(2m)$ is the noninteracting Fermi energy and $\chi(q,z)$, with $z=\omega\, (i\omega)$ is the linear density-density response function along the real (imaginary) frequency axis, which could be obtained from
	\begin{equation}\label{response}
	\chi (q,z)=\frac{\chi_{0} (q,z)}{1-w(q)\chi_{0} (q,z)},
	\end{equation} 
	in terms of the noninteracting density-density response function $\chi_{0} (q,z)$, whose analytic form is given in Appendix \ref{sec:app1}.
	
	\section{Quasiparticle energy and lifetime}
	The interaction between particles has two main effects on the energy of quasiparticles. First, the energy dispersion relation and the effective mass of QPs are renormalized which are determined by the real part of the self-energy. Second,  owing to the inelastic scatterings, the quasiparticles acquire a finite decay rate proportional to the imaginary part of the self-energy.
	\begin{figure}
		\centering
		\includegraphics[width=0.95\linewidth]{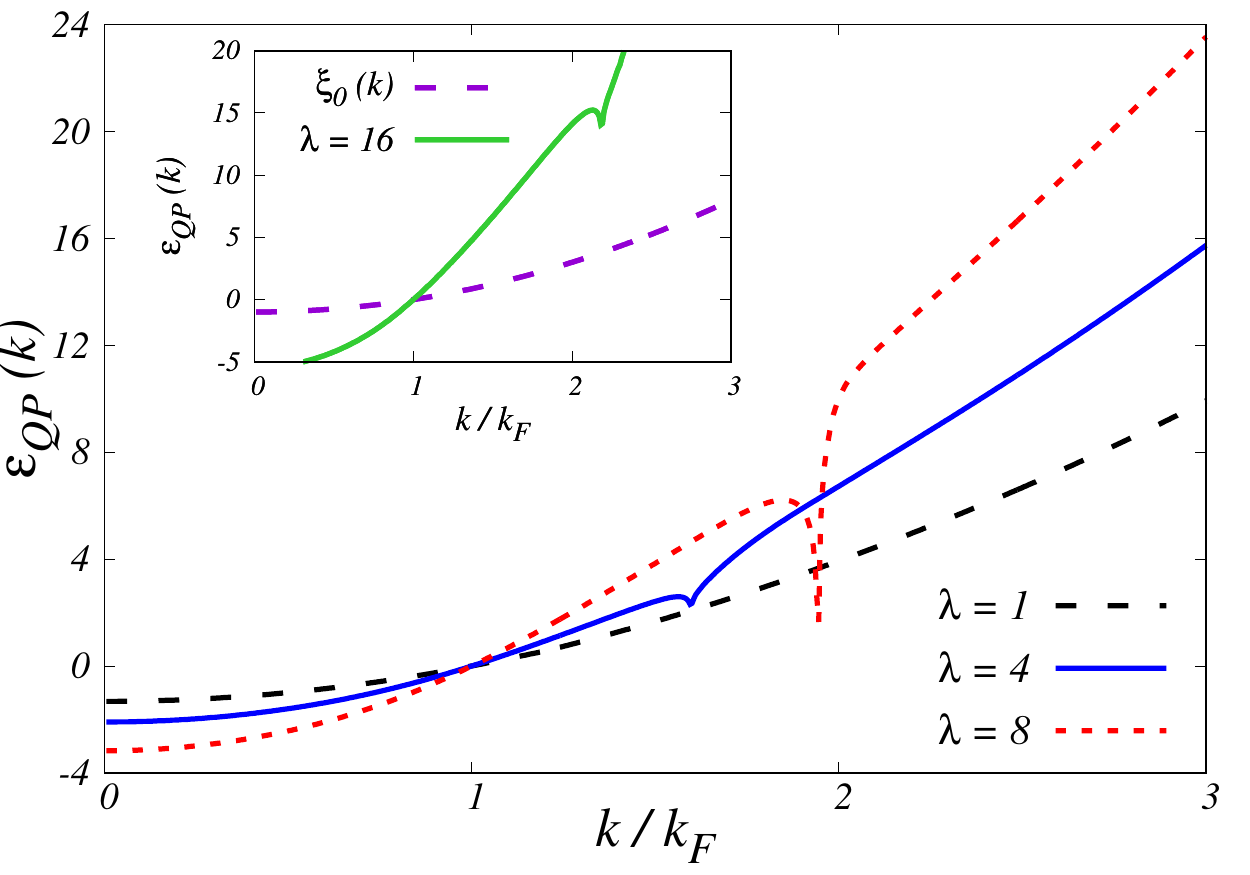} \\
		\includegraphics[width=0.95\linewidth]{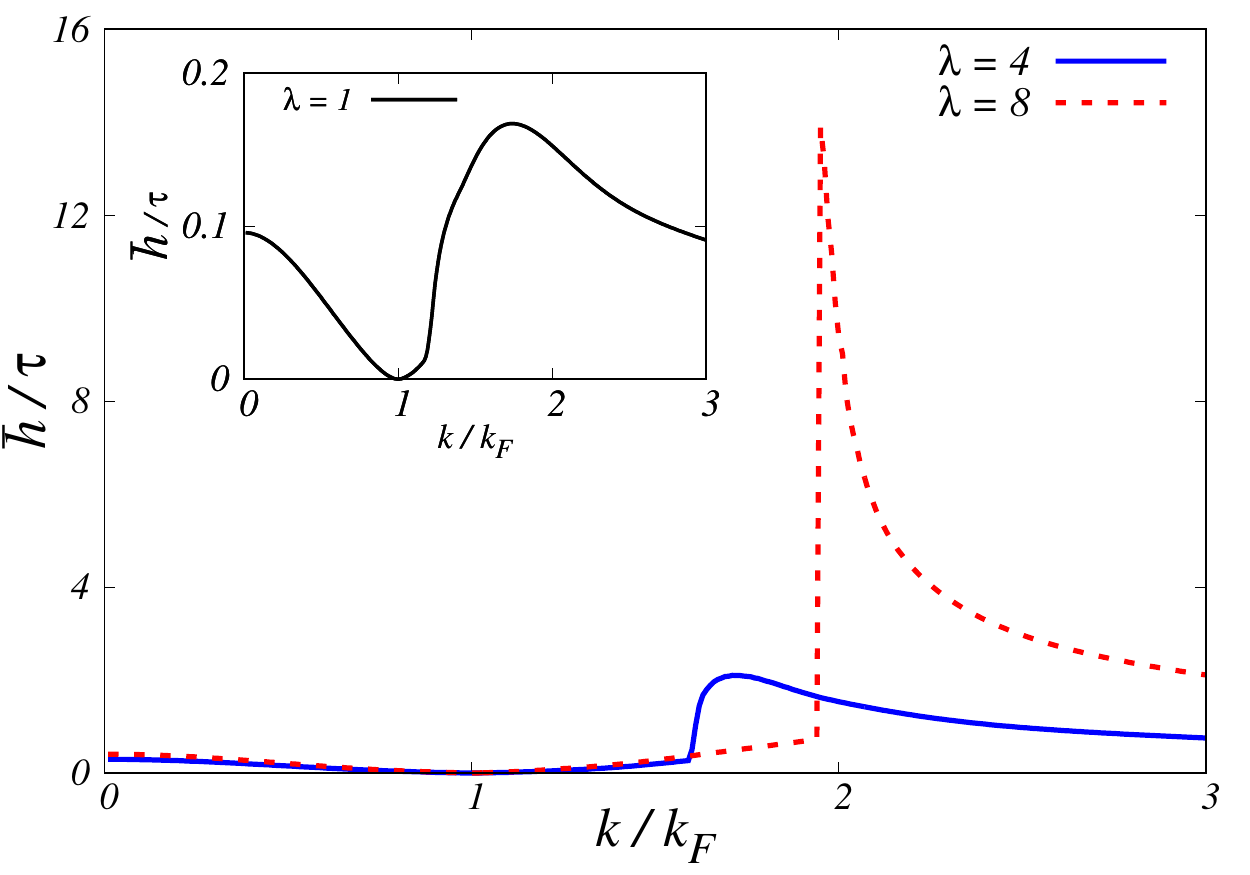}
		\caption{Top panel: The quasiparticle energy (in units of $\varepsilon _{\rm F}$) as a function of $k$ at various coupling strengths. The inset compares the quasiparticle energy of strongly coupled dipolar Fermi liquid at  $\lambda=16$ with the energy dispersion of a noninteracting system. Bottom panel: The quasiparticle decay rate $\hbar/\tau$ (in units of $\varepsilon _{\rm F}$) as a function of $k$ for $\lambda=4$ and $\lambda=8$. The quasiparticle decay rate, $\Im m\, \Sigma ( k, \xi_0 ( k))$, vanishes at the Fermi level as $|k - k_{\rm F}|^2$ which is one of the main features of the Landau theory of Fermi liquid~\cite{Vignale}. The inset of bottom panel shows the decay rate at weak couplings (i.e., $\lambda=1$), where the jump at $k_c$ is absent.
			\label{fig:E}}
	\end{figure}
	
	The excitation spectrum of quasiparticles, measured with respect to the interacting chemical potential, could be written as
	$\varepsilon_{\rm QP}(k) = \xi_0(k)+ \Re e\, {\tilde \Sigma}(k,E)|_{E = \varepsilon_{\rm QP}(k)}$,
	where $\Re e \, {\tilde \Sigma}(k, E) = \Re e\, \Sigma (k, E) - \Re e\, \Sigma (k_{\rm F} , 0)$.
	In principle, the equation of the excitation spectrum should be solved self-consistently.
	However, if the interaction is not too strong one can resort to the ``on-shell approximation" (OSA), replacing $\varepsilon_{\rm QP}(k) \rightarrow \xi_0(k)$ in the argument of self-energy~\cite{Asgari1,Asgari2,Vignale} and thus find
	\begin{equation}\label{OSA-energy}
	\varepsilon^{\rm OSA}_{\rm QP}(k) \approx \xi_0(k) + \Re e \,{\tilde \Sigma}(k,\xi_0(k)).
	\end{equation}
	Figure\,\ref{fig:E} shows the wave vector dependance of quasiparticle energy and the inverse of its lifetime
	$\hbar/\tau = 2 |\Im m\, \Sigma ( k , \xi_0 ( k))|$, obtained within the on-shell approximation for self-energy at different values of the coupling strength $\lambda$.
	Generally, two intrinsic mechanisms contribute to the scattering of quasiparticles: i) the excitation of particle-hole pairs which is dominated at long wavelengths, and ii) the excitation of zero-sound collective mode that turns on at a threshold wave vector $k_c$~\cite{Qaiumzadeh, Jalabert}.
	At weak couplings, the system behaves qualitatively similar to an ideal Fermi gas thanks to the cancellation between static HF and dynamical contributions to the self-energy. At intermediate and strong coupling strengths, for $k < k_{\rm F}$ the QP is only moderately affected by the many-body effects, but for $k > k_{\rm F}$ and especially around $k_{\rm c}$, the QP spectrum is strongly affected as the quasiparticle energy loses most of its energy through inelastic scattering with collective modes. A strong dip in the QP spectrum is the zero-sound dip and its position moves to higher wave vectors increasing the coupling strength. This zero-sound dip, which resembles the maxon-like dip reported in 2D $^3$He~\cite{Boronat} and the plasmon dip in 2D electron liquids~\cite{Asgari1, Asgari2}, originates from the decay of particle-hole pairs into collective modes with conserved momentum and energy~\cite{Jalabert, Vignale}.
	At intermediate and strong couplings a finite jump in the decay rate takes place at $k_c$, as the scattering rate is drastically increased at the zero-sound dip. A similar jump has been reported in a 2D electron liquid~\cite{Asgari1} in which it was associated with the non-zero oscillator strength of the plasmon-pole at the wave vector of plasmon dip.
The quasiparticle decay rate vanishes as  $\approx (E -\mu)^{\beta}$ for $E \rightarrow \mu $ at $k \rightarrow k_{F}$. 
	From our numerical results, we find that at weak and intermediate couplings $\beta \lesssim 2$, while at strong couplings $\beta$ is slightly smaller than $2$, but we still get $1 < \beta \leq 2$ (see, Appendix \ref{sec:app2} for more details). This is one of the main features of the Landau theory for the Fermi liquid~\cite{Vignale}.

\section{Renormalization constant and effective mass}
In the presence of interactions, discontinuity of the momentum distribution at $k =  k_{\rm F}$, that is measured by the renormalization constant
\be
Z =1/\left[1-\partial_{E}\Re e \, \Sigma (k,E)|_{k = k_{\rm F} , E = 0}\right],
\ee
	is less than unity~\cite{Vignale}.
	The many-body effective mass at the Fermi level could be obtained from the slope of interacting excitation spectrum~\cite{Vignale} 
	$\hbar^{2}k_{\rm F}/m^*=\left.\mathrm{d}\varepsilon_{\rm QP}(k)/\mathrm{d}k\right|_{k\to k_{\rm F}}$.
	Depending on whether the quasiparticle energy is calculated by solving the self-consistent Dyson equation or by using the OSA, we would find different results for the effective mass.
	The effective mass within the Dyson approximation $m^*_{\rm D}$ is given by
	\begin{equation}\label{m_D}
	\frac{m}{m^*_{\rm D}}=Z\left[1+ (m/\hbar^{2}k_{\rm F}) \partial_{k} \Re e \Sigma (k,E)|_{k = k_{\rm F} , E = 0}\right] ,
	\end{equation}
	and the OSA for the QP energy for the many-body effective mass gives
	\be\label{m_OSA}
	\begin{split}
		\frac{m}{m^*_{\rm OSA}}=1+&(m/\hbar^2k_{\rm F}) \partial_{k} \Re e \Sigma (k,E)|_{k = k_{\rm F} , E = 0} \\
		& + \partial_{E} \Re e \Sigma (k,E)|_{k = k_{\rm F} , E = 0} .
	\end{split}
	\ee
	\begin{figure}
		\centering
		\includegraphics[width=1.0\linewidth]{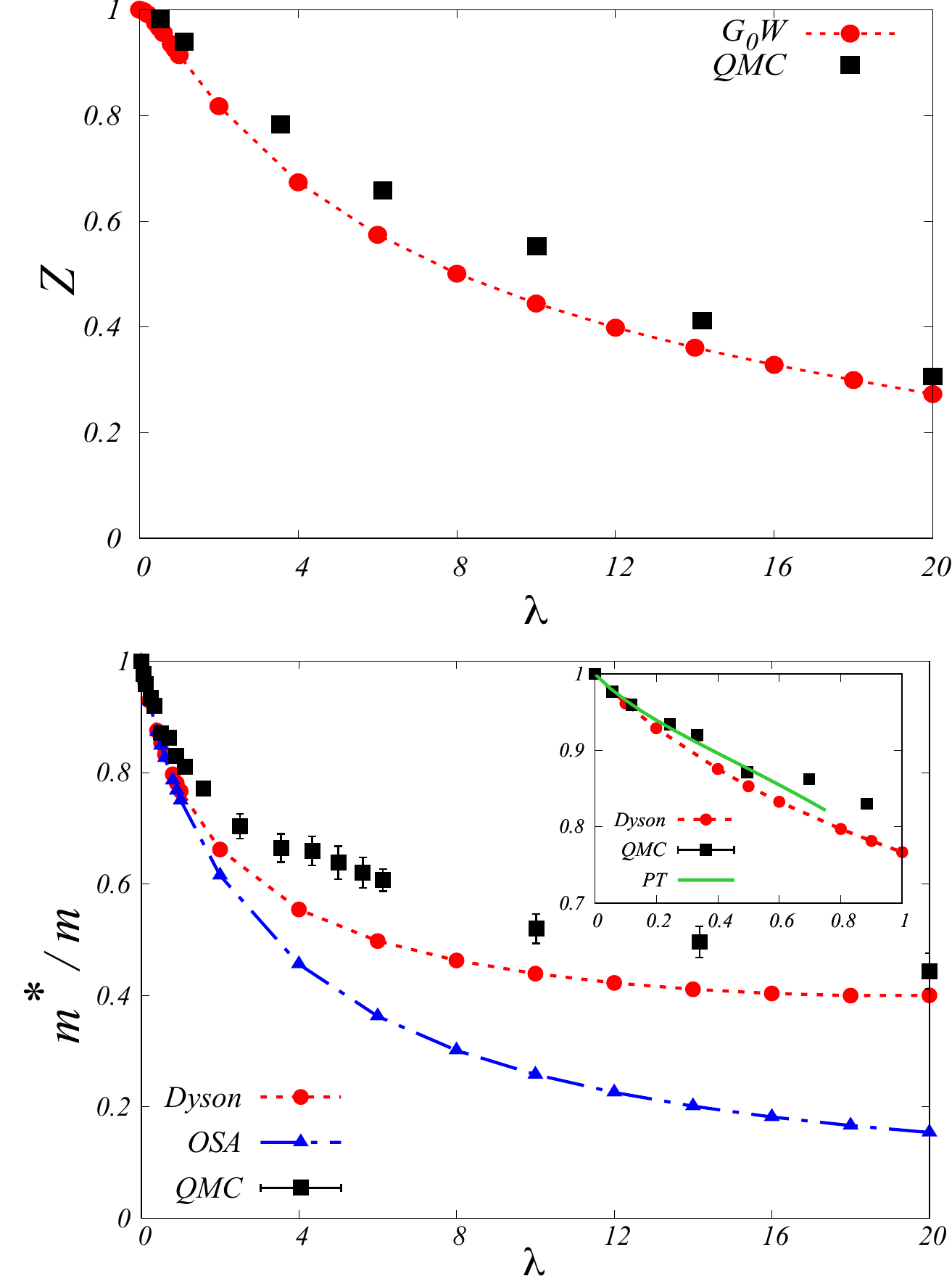}
		\caption{Top panel: The renormalization constant $Z$ of a 2D dipolar Fermi liquid as a function of the dimensionless coupling constant $\lambda$, calculated within the $G_0W$ approximation and compared with QMC results of Ref.\,\cite{Matveeva}.
Bottom panel: The relative effective mass of a 2D dipolar Fermi liquid $m^*/m$ as a function of the dimensionless coupling constant $\lambda$, calculated within the $G_0W$ approximation and with Dyson and on-shell approximations for the self-energy, obtained from Eqs.\,\eqref{m_D} and \eqref{m_OSA}, respectively. The results are compared with the QMC data of Ref.\,\cite{Matveeva}.
The inset compares our Dyson results for the relative effective mass with the QMC data of Ref.\,\cite{Matveeva} and the analytic second-order perturbation theory results of Ref.\,\cite{Lu1} at weak couplings.
\label{fig:mass}}
\end{figure}
In Fig.\,\ref{fig:mass} we compare our numerical results for the renormalization constant $Z$ and the effective mass $m^*$ with the QMC results of Matveeva and Giorgini~\cite{Matveeva} over a wide range of coupling strengths.
In the weak coupling regime, the self-energy is dominated by the direct and exchange effects. As the Hartree-Fock self-energy is static the renormalization constant remains close to one at $\lambda <1$. As the coupling constant increases, the effects of correlation become important which causes the reduction of the renormalization constant. A strong suppression of the renormalization constant is visible at strong coupling strengths, but $Z$ never reaches zero. As the Landau Fermi liquid theory implies $0< Z \leq 1$~\cite{Vignale}, this could be taken as an indication of the Fermi liquid picture remaining valid for dipolar Fermi system up to very strong couplings. Interestingly, our $G_0W$ results agree very well with the QMC data of Matveeva and Giorgini~\cite{Matveeva} over the whole range of coupling strengths where the homogeneous liquid phase is predicted to be stable. We should note that from the QMC simulation~\cite{Matveeva} phase transition to Wigner crystal is expected at $\lambda=25\pm3$.
 
Both OSA and Dyson methods predict a strong reduction of effective mass with respect to its bare value at strong couplings, but the Dyson results are in a much better agreement with the QMC data~\cite{Matveeva}.
In the inset of  the bottom panel in Fig.\,\ref{fig:mass} we compare our Dyson results for the effective mass at weak couplings, with the second order perturbative expansion results of Liu and Shlyapnikov~\cite{Lu1}
$m/m^*=1+4\lambda/(3\pi)+\lambda^2\ln(0.65\lambda)/4$. Note that the second term on the right-hand side of this expression is the HF contribution to the renormalization of effective mass and the third term is the leading order contribution beyond the HF approximation.
	
The Hartree-Fock contribution to the self-energy $\Sigma_{\rm HF}(k)$ has a positive slope at the Fermi wave vector and hence decreases the effective mass, whereas the dynamic contribution to the self-energy $\Sigma_{\rm \rho}(k)$ has a negative slope and tends to enhance $m^*$~\cite{Seydi}.  But the Hartree-Fock contribution dominates over the dynamical term over the whole range of coupling strengths and the effective mass is suppressed with respect to its noninteracting value.
\section{Dynamical structure factor and spectral function}
\begin{figure}
\centering
\includegraphics[width=1.0\linewidth]{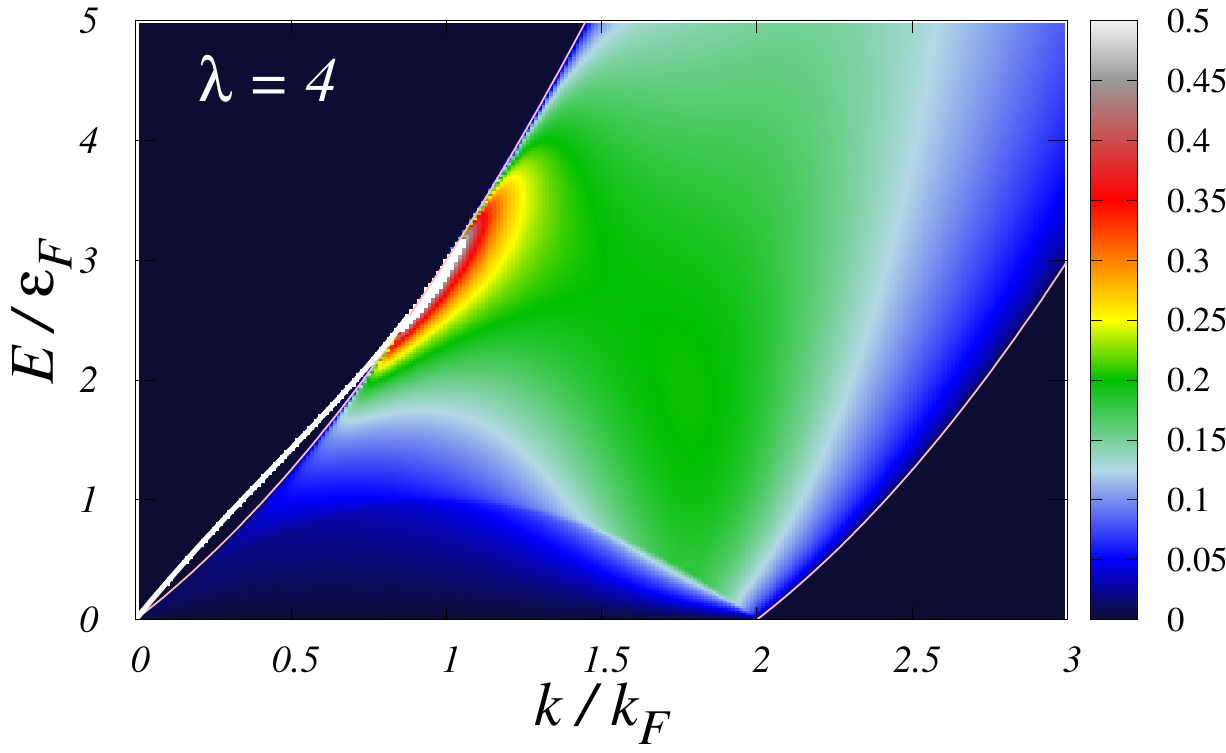}\\
\includegraphics[width=1.0\linewidth]{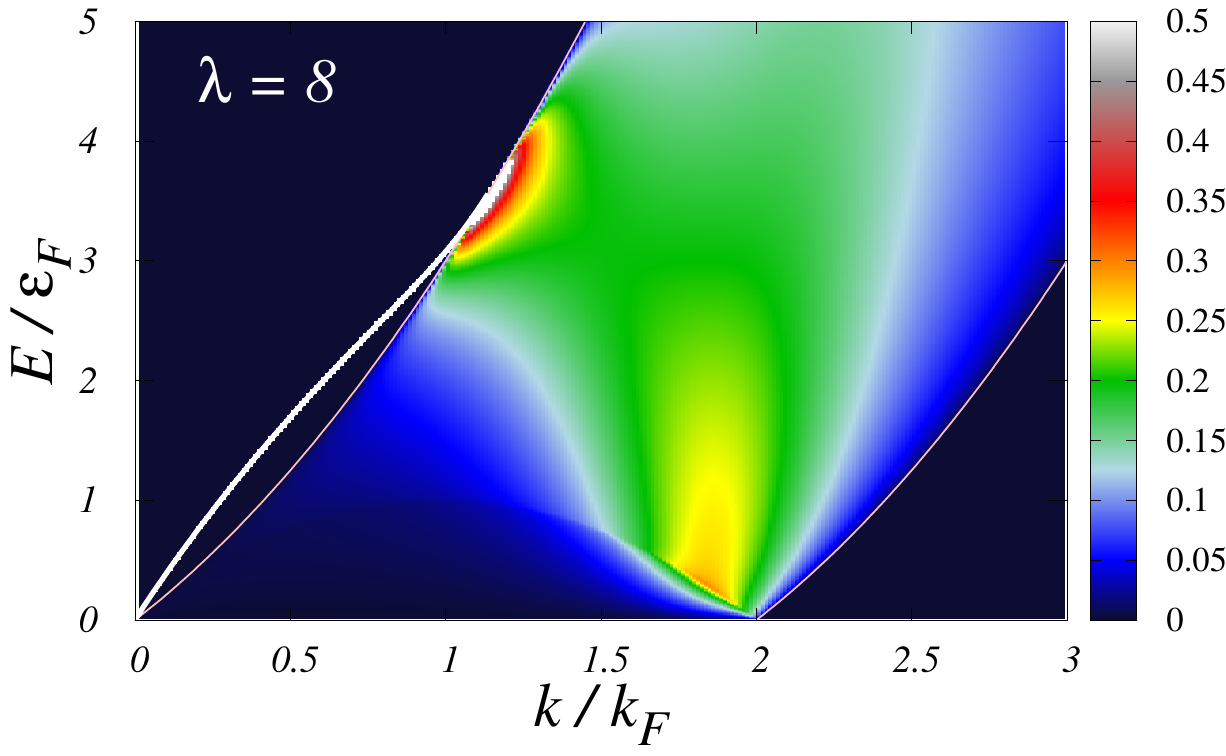} 
\caption{Density plots of the dynamical structure factor of a 2D dipolar Fermi liquid [in units of $\hbar/(\pi \varepsilon_{\rm F})$] as a functional of energy $E$ and wave vector $k$ calculated for different values of the coupling constant $\lambda$. Note that the Dirac delta-function form of the structure factor along the dispersion of zero-sound mode outside the single particle continuum has been broadened by $10^{-4}\,\varepsilon_F$  for better visibility.
\label{fig:Skw}}
\end{figure}
The dynamical structure factor at the zero temperature is related to the imaginary part of the density-density response function as~\cite{Vignale}
\be
S(k,\omega)=-\frac{\hbar}{n\pi}\Im m \chi(k,\omega),
\ee
where the interacting density-density response function $\chi(k,\omega)$ is approximated by Eq.\,\eqref{response}. The behavior of $S(k,\omega)$ for different coupling constants have been illustrated in Fig.\,\ref{fig:Skw} as the density plots in the energy-wave vector plane. The regions of particle-hole continuum together with the dispersions of zero-sound modes are clearly visible. Note that as the dynamical behavior of the effective interaction $w(q,\omega)$ is not included in our formalism, the imaginary part of the density-density response function remains zero outside the particle-hole continuum, except on top of the zero-sound dispersion where it acquires a Dirac delta function form~\cite{Vignale}. 
The dispersion of collective mode enters the particle-hole continuum at a characteristic wave vector $k_{\rm zs}$ (e.g., $k_{\rm zs}\approx 0.8\,k_{\rm F}$ for $\lambda=4$ and $k_{\rm zs}\approx k_{\rm F}$ for $\lambda=8$) after which the Landau damping of the collective mode begins.
\begin{figure*}
\centering
\includegraphics[width=1.0\textwidth]{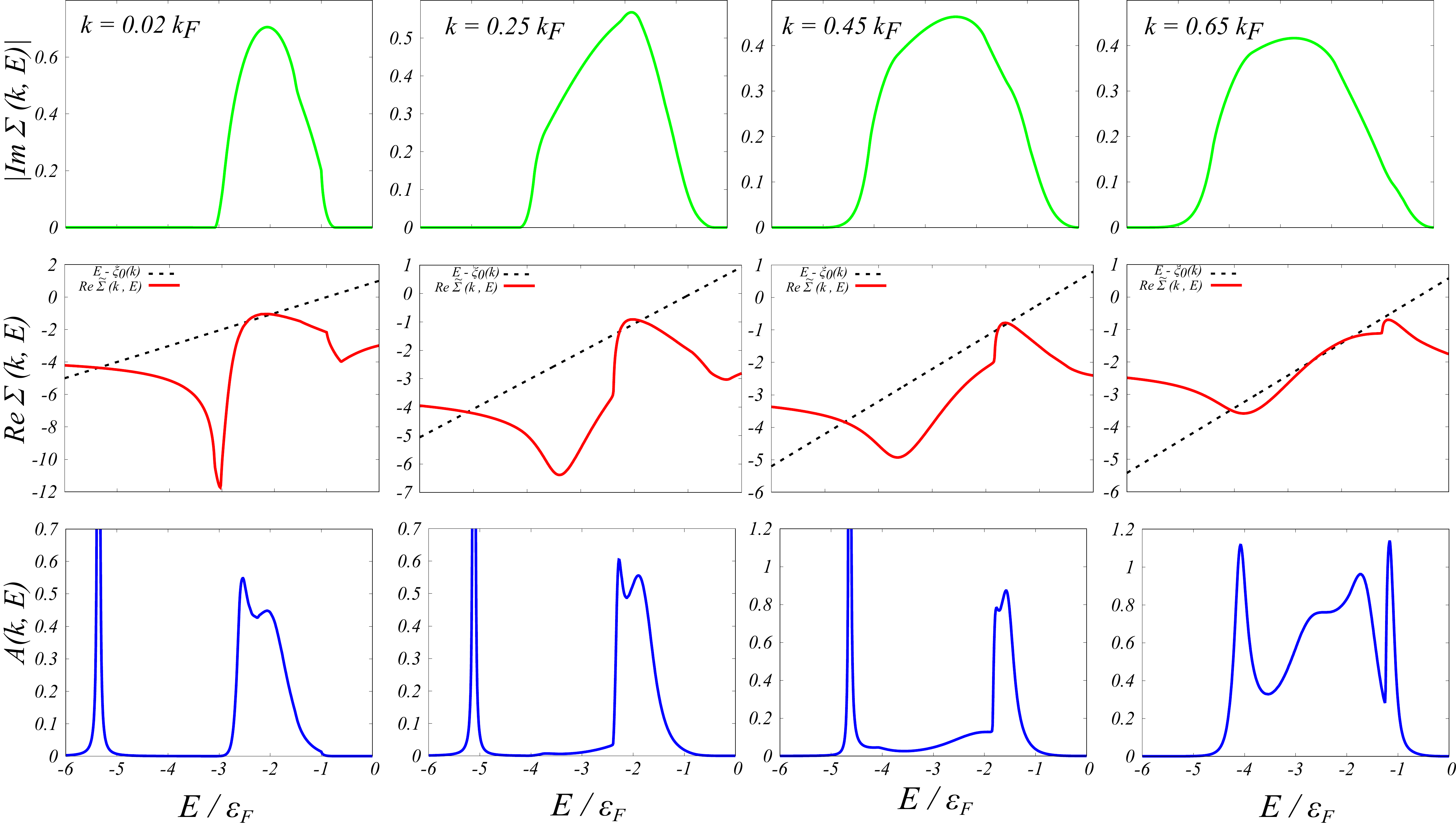}
\caption{The imaginary (top row) and real (middle row) parts of the $G_0W$ self-energy of a 2D dipolar Fermi liquid (both in the units of $\varepsilon_{\rm F}$) as functions of energy at four fixed values of the wave vectors $k = 0.02\, k_{\rm F}, 0.25\, k_{\rm F}, 0.45\, k_{\rm F}$, and $0.65\, k_{\rm F}$ (from left to right). In the middle row the straight dashed lines show $E-\xi_0(k)$ whose cross-section with the real part of the self-energy indicate the positions of interacting quasiparticle energies. 
In the bottom row, the spectral functions (in the units of $1/\varepsilon_{\rm F}$) are shown as functions of energy at the same fixed values of the wave vector as upper panels. The positions of peaks in the spectral function correspond to the crossing points of two curves in the middle row and their width is specified by the imaginary part of the self-energy (top row). The undamped peaks have been broadened by $0.005 \,\varepsilon_{\rm F}$ for better visibility.
\label{fig:8FHNC}}
\end{figure*} 

\begin{figure}\label{fig:Ak0}
\centering
\includegraphics[width=1.0\linewidth]{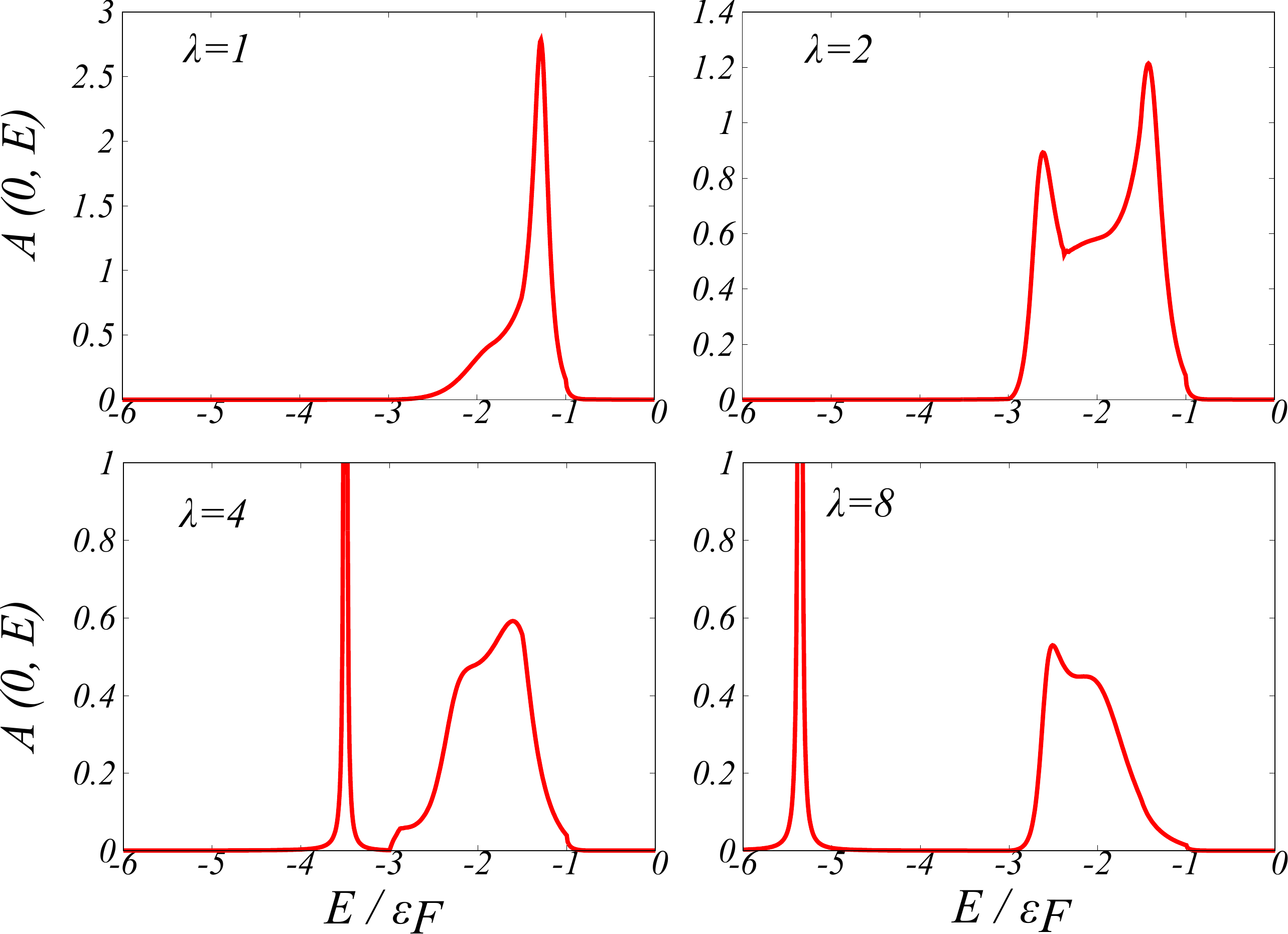} 	
		\caption{The spectral function  of a 2D dipolar Fermi liquid (in units of $1 / {\varepsilon_{\rm F}}$) versus energy, calculated at $k = 0$ for different values of the coupling constant. The undamped peaks have been broadened by $0.005 \,\varepsilon_{\rm F}$ for better visibility.
	\label{fig:A0}
}
\end{figure} 
	\begin{figure*}
		\centering
		\begin{tabular}{ccc}
			\includegraphics[width=0.35\linewidth]{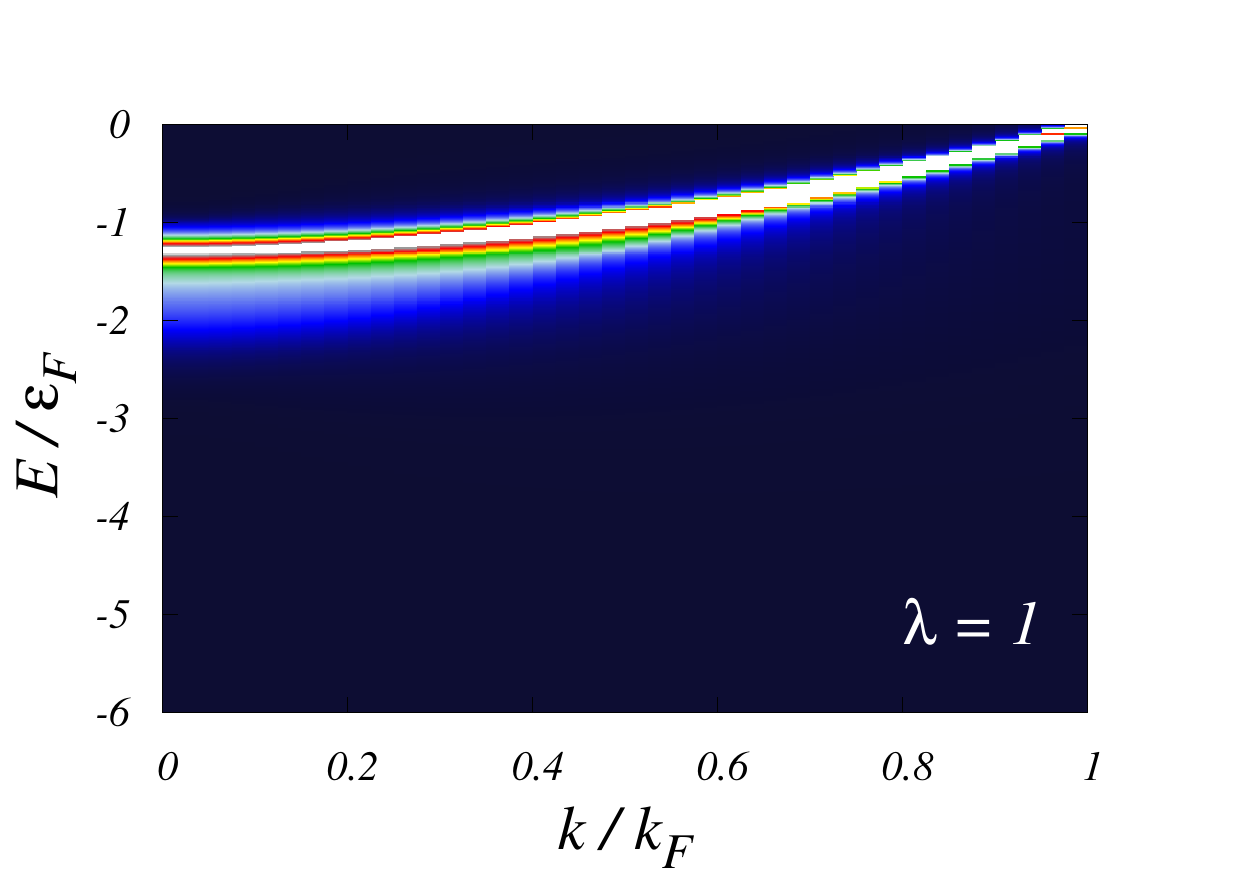}&
			 \includegraphics[width=0.35\linewidth]{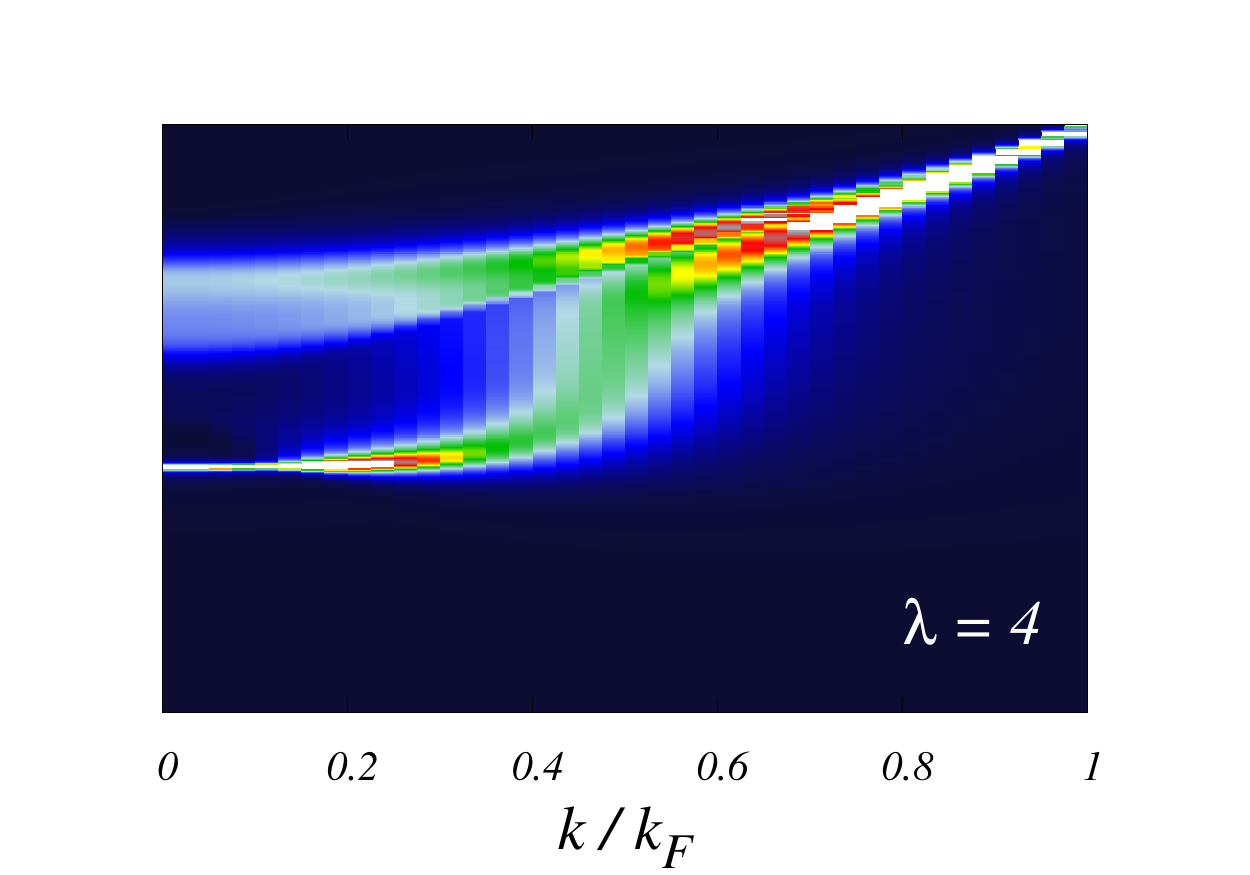}&
			  \includegraphics[width=0.35\linewidth]{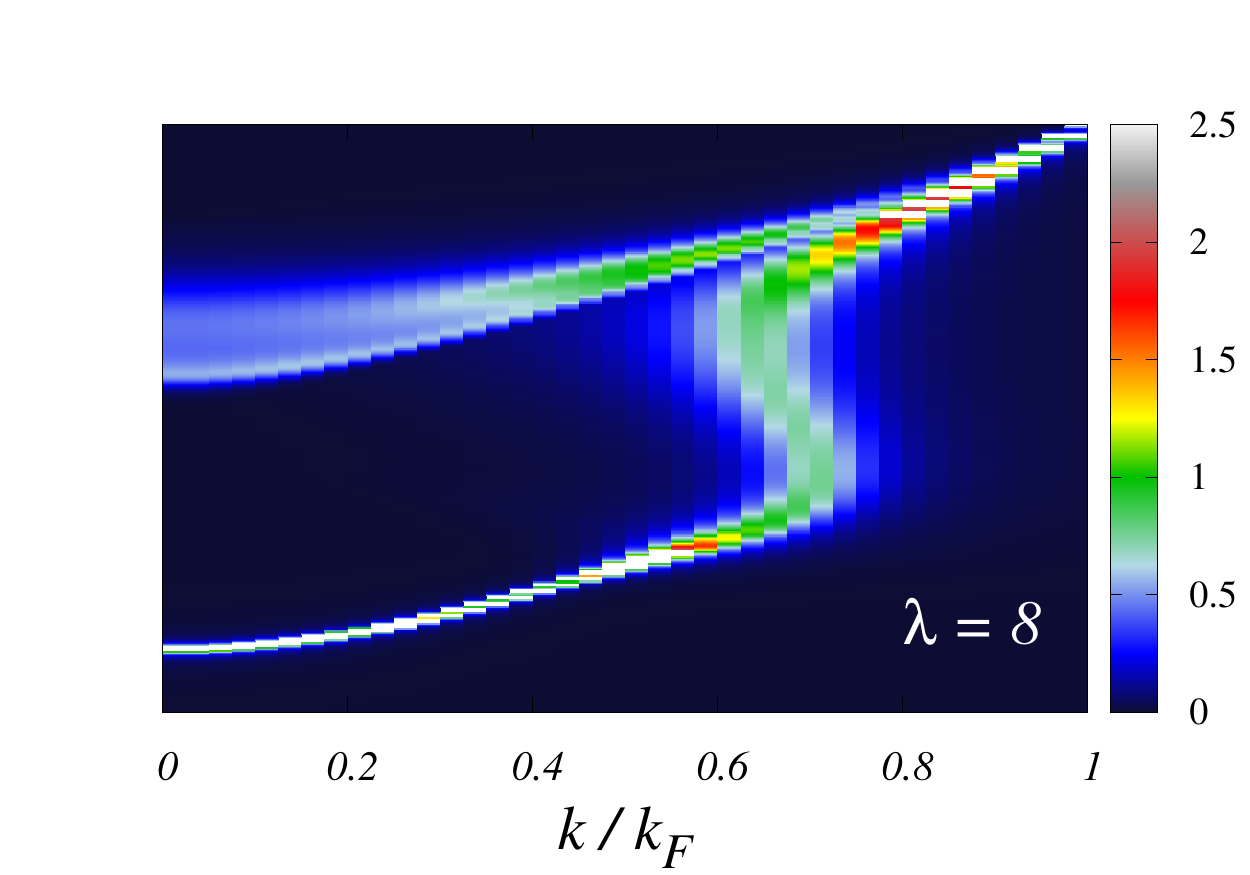}
		\end{tabular}
		\caption{(Color online) Density plot of the spectral function $A (k , E)$ (in units of $1/\varepsilon_{\rm F}$) as a function of energy $E$ and wave vector $k$, obtained from Eq.\,\eqref{eq:A_kw} for three different values of the coupling constants $\lambda = 1$ (left), $\lambda = 4$ (middle), and $\lambda = 8$ (right). The spectral function has been broadened by $0.005\,\varepsilon_{\rm F}$ for a better visibility of the undamped delta-like peaks.\label{fig:Akw}}
	\end{figure*}
	
	The spectral function $A(k, E)$ which is a measure of having a particle with momentum $k$ and energy $E$, can be obtained from the imaginary part of the retarded Green's function. For a noninteracting system, the spectral function has a delta-function form. In the presence of interactions, the modification of single particle Green's function $G^{-1}(k,E) = G^{-1}_0 (k,E) - \Sigma (k,E)$ usually broadens the spectral function $A(k,E) = -\Im m \, G(k,E)/\pi$.
	The interacting spectral function in terms of the self-energy and the noninteracting energy dispersion $\xi_0(k)$ is written as~\cite{Mahan,Vignale}
	\begin{equation}\label{eq:A_kw}
	A(k, E) = \frac{-\Im m \Sigma (k , E)/\pi }
	{[E - \xi_0(k) - \Re e {\tilde \Sigma} (k , E)]^2 +[\Im m \Sigma (k , E)]^2}.
	\end{equation}
	The spectral function is a positive-definite quantity and its exact expression should satisfy the sum-rule $\int_{-\infty}^{\infty} \mathrm{d} E A(k , E)  = 1$.
	A fraction $Z$ of the total spectral weight is absorbed by the quasiparticle peak at $E=\varepsilon_{\rm QP}(k)$, and the remaining $1-Z$ weight is distributed over the background~\cite{Vignale}.
	
	Vanishing of the expression inside the first square bracket in the denominator of Eq.\,\eqref{eq:A_kw} manifests itself as a resonance in the spectral function.
	These resonances represent quasiparticles with specific energies and possible finite lifetimes.
	The position of peaks in the single-particle spectral function could be obtained from the solution of Dyson's equation for quasiparticle energy. 
	Figure\,\ref{fig:8FHNC} illustrates the typical behavior of the imaginary and real parts of the self-energy as well as the spectral function $A(k,E)$ of the dipolar Fermi gas for $\lambda = 8$. The solutions of the Dyson equation are obtained from the intersections of the real part of the self-energy $ \Re e\, \tilde{\Sigma} (k, E)$ and the straight line $E - \xi_0(k)$. 
	Three peaks in the spectral function correspond to three distinct solutions of the Dyson equation for $\lambda=8$.   
	If we account for these peaks according to decreasing energy, the first peak is associated with the quasiparticle solution which is shifted from the noninteracting energy and has a finite width corresponding to the non-zero damping rate. The second and third peaks result from the coupling of quasiparticles and collective (i.e., zero-sound) modes. As it is evident from the behavior of the imaginary part of the self-energy, the composite quasiparticles are undamped in the long wavelength limit.
	Approximately for $k \gtrsim 0.45 k_{\rm F}$, the third peak acquires a finite width and becomes damped and for $k \gtrsim 0.7 k_{\rm F}$ only the quasiparticle excitation peak survives in the spectral function.
	
	It is convenient to look at the behavior of $\Im m \Sigma (k , E)$ at $k=0$. In this limit, the initial QP energy is $E_i=E+\varepsilon_{\rm F}$ and the final energy is $E_f=\xi_0(q)+\varepsilon_{\rm F}$. When the difference is equal to the mode energy, $\hbar \omega$ the initial QP can decay by emitting a composite \emph{particle-zero sound} excitation. Since $\hbar \omega \geq \xi_0(q)$ and $E_f-E_i \leq \xi_0(q)$, therefore an initial QP with $E_i<0$ can decay into a final state through a resonance process in which $E_f>0$ when $d \omega/dq=\hbar^{-1}d\xi_0/dq=\hbar q/2m$. When these conditions are met, $\Im m \Sigma (k=0 , E)$ peaks at a specific $E$ and its Kramers-Kronig transformation $\Re e  \Sigma (0 , E)$ changes sign rapidly around that energy.
	In Fig.\,\ref{fig:A0}, we illustrate the behavior of spectral function at $k = 0 $ and for different values of the coupling constant. As it is seen the composite quasiparticle peak emerges at $\lambda \approx 2$.
	Increasing the coupling strength the composite peak moves towards lower energies and becomes well separated from the quasiparticle excitation peak.
	
	In Fig.\,\ref{fig:Akw} we show the density plot of the spectral function at three different values of the coupling constant $\lambda = 1,4$, and $8$. In the weak coupling regime (cf., the left panel in Fig.\,\ref{fig:Akw}) only the particle-hole excitation dispersion is visible in the spectral function. In addition to the QP peak, the composite \emph{quasiparticle-zero sound} excitation peak emerges around $\lambda\approx 2$ by illustrating a peak in the $\Im m \Sigma (k=0 , E)$ around $E\approx - 2.7 \varepsilon_{\rm F}$. 
	This new peak corresponds to the bound states of quasiparticles with the zero-sound mode in the 2D dipolar Fermi liquid. Owing to the repulsion between QP and composite QP resonances, a gap-like feature between these two bands are visible at long wavelengths.
	The composite quasiparticles are undamped at small $k$ and their dispersion eventually merges with the dispersion of QPs at a characteristic $\lambda$-dependent wave vector. This corresponds to the wave vector where the dispersion of zero-sound mode enters the particle-hole continuum and gets Landau-damped. 
	The emergence of a similar feature in electron liquids which is called plasmaron has been first proposed by Lundqvist~\cite{Lundqvist}.
	The plasmaron peak has been predicted in 2D electron liquid~\cite{Jalabert} and graphene ~\cite{Polini, Hwang} as well and has been experimentally verified in doped graphene~\cite{Bostwick} through the angle-resolved photoemission spectroscopy.
	But a similar feature has never been observed in neutral Fermi liquids.

	\section{summary and conclusion} 
	We explore the quasiparticle properties of a spin-polarized 2D DFL. We employ the $G_0W$ approximation with an effective interaction extracted from the very accurate FHNC data for the interacting static structure factor. With such an effective interaction, we are able to achieve a very good agreement with the QMC data for effective mass and renormalization constant up to very strong coupling regimes.
	We should here note that based on the existing experiences with the electronic systems~\cite{Asgari1, Asgari2} results of the $G_0W$ method for the effective mass is very sensitive to the approximations one adopts for the effective interaction and it is generally very difficult to obtain a very good agreement with QMC data over a large range of coupling strengths. 
	Our findings suggest that a modified $G_0W$ approach armed with an effective particle-particle interaction extracted from accurate static structure factor data might perform equally well for other strongly interacting quantum fluids too. A similar procedure could be also implemented in \emph{ab-initio} electronic structure packages to improve the $GW$-DFT results for the quasiparticle spectrum.
	Then, we switch to the investigation of the spectral function below the Fermi energy. Apart from the conventional quasiparticle dispersion, we observe a novel composite \emph{particle-zero sound} dispersion at intermediate and strong couplings which is a bound state of a dipolar hole below the Fermi level and the collective density oscillation.
	This massive mode is undamped at long wavelengths and we would expect it to be observable through RF spectroscopy of ultracold dipolar systems consisting of non-reactive fermionic polar molecules at intermediate and high densities. With polar molecules whose dipolar length $r_0$ could easily reach thousands of nanometers, an average planar density of $10^6-10^7$ cm$^{-2}$ would suffice to observe the novel features in the spectral function predicted here. 
	
\acknowledgements
	We are grateful to G. V. Shlyapnikov, G. Baskaran, and M. R. Bakhtiari for very helpful discussions. BT is supported by TUBA and TUBITAK. 
	
	\appendix
	\section{Density-density response function of a noninteracting 2D system}\label{sec:app1}
	The density-density response function of a noninteracting spin-polarized two-dimensional Fermi system, along the real frequency axis is the famous Stern-Lindhard function~\cite{Vignale}
	\begin{equation}\label{eq: chi_0(q,w)}
	\chi_{\rm 0}(q , \omega) = \frac{1}{S}
	\sum_{\bf k} \frac{n_{\rm FD}[\varepsilon_0({\bf k})] - n_{\rm FD}[\varepsilon_0({ \bf k +\bf q})] }{\hbar\omega + \varepsilon_0({\bf k})-  \varepsilon_0({ \bf k +\bf q})+i \eta},
	\end{equation}
	where $S$ is the sample area and $\eta$ is an infinitesimal positive quantity.
	After performing the sum over $\kv$, the real and the imaginary parts of the response function read
	\begin{widetext}\label{Re_chi0}
		\begin{equation}
		\Re e\, \chi_{\rm 0}(q , \omega) = -\nu_{0}\left\{ 1 +  \frac{1}{\tilde q}\left[{\rm sgn}(\nu_{-}) \Theta(\nu_{-}^2 -1)\sqrt{\nu_{ -}^2 -1} - {\rm sgn}(\nu_{ +}) \Theta(\nu_{ +}^2 -1)\sqrt{\nu_{ +}^2 -1}\right]\right\}, 
		\end{equation}
	\end{widetext}
	and
	\begin{widetext}
		\begin{equation}
		\Im m \, \chi_{\rm 0}(q , \omega) = -\frac{\nu_{ 0}}{\tilde q}\left[\Theta(\nu_{\rm -}^2 -1)\sqrt{\nu_{\rm -}^2 -1} - \Theta(\nu_{\rm +}^2 -1)\sqrt{\nu_{\rm +}^2 -1}\right]. 
		\end{equation}
	\end{widetext}
	Here, $\nu_0=m/(2\pi\hbar^2)$ is the density of states per unit area of a spin-polarized 2D system and $\nu_{ \pm} = \tilde{\omega}/(2\tilde q) \pm {\tilde q}/2$ with ${\tilde q}=q/k_{\rm F}$, and ${\tilde \omega}=\hbar\omega/(2 \varepsilon_{\rm F})$. 
	The Lindhard function along the imaginary frequency axis, after some straightforward algebra reads
	\begin{equation}\label{chi0}
	\chi_{\rm 0}(q,i\omega) = -\nu_{\rm 0}\left(1-\frac{\sqrt{2}}{\tilde q}\sqrt{a+\sqrt{a^{2}+{\tilde \omega}^2}}\right),
	\end{equation}
	where we have defined $ a={\tilde q}^2/4-{\tilde \omega}^2/{\tilde q}^2-1$.
	
\section{The imaginary part of the self-energy}\label{sec:app2}
	The only contribution to the imaginary part of the self-energy arises from the imaginary part of the pole term
	\begin{widetext}
		\begin{equation}
		\Im m \, \Sigma (k , E) = \int \frac{\mathrm{d}^2{\bf q}}{(2\pi)^2} w^{2}(q) \Im m \, \chi(q,E-\xi_0 (\textbf{k} - \textbf{q}))\left[\Theta(E-\xi_0(\textbf{k} - \textbf{q}))-\Theta(-\xi_0(\textbf{k} - \textbf{q})\right],
		\end{equation}
	\end{widetext}
	where
	\begin{widetext}
		\begin{equation}
		\Im m \, \chi(q,\omega) = \frac{\Im m \, \chi_{0}(q,\omega)}{\left[1-w(q) \Re e \ \chi_{0}(q,\omega)\right]^2 + \left[w(q)\Im\, m \ \chi_{0}(q,\omega)\right]^2}
		- \frac{\pi \delta(\omega - \Omega_{ZS}(q)) \Re e \, \chi_{0}(q , \Omega_{ZS}(q))}
		{w(q)\arrowvert \partial_\omega \Re e\, \chi_{0}(q , \omega) \arrowvert_{\omega = \Omega_{ZS}(q)}}.
		\end{equation}
	\end{widetext}
	The second term on the right-hand-side of this expression is the collective mode contribution to the imaginary part of the response function, with $\Omega_{ZS}(q)$ being the frequency of zero sound at wave-vector $q$. By defining the dimensionless parameters $\tilde{k} = k / k_{\rm F}$, $\tilde{q} = q / k_{\rm F}$, $\tilde{E} = E / \varepsilon_{\rm F}$, and $ y = \tilde{E} -\tilde{k}^2 -\tilde{q}^2 + 2\tilde{k}\tilde{q}\cos \phi + 1$, we find
	\begin{widetext}
		\be
		\begin{split}
			&\Im m \, \Sigma (k , E) =
			-2  \varepsilon_{\rm F}\int_{0}^{\infty} \mathrm{d}{\tilde {q}} \tilde {q}w(\tilde{q}) \int _{y_{1}}^{y_{2}}\mathrm{d}y \frac{ \delta(y - \tilde{\Omega}_{ZS}) \Re e \, \chi_{0}(\tilde{q} ,  \tilde{\Omega}_{ZS})}{w(\tilde{q})\arrowvert \partial_y \Re  e\, \chi_{0}(\tilde{q} , y)\arrowvert_{y =  \tilde{\Omega}_{ZS}}} \left[\Theta(y)-\Theta(y-\tilde{E})\right]  \\
			&+ \frac{2}{\pi} \varepsilon_{\rm F}\int_{0}^{\infty} \mathrm{d}{\tilde {q}} \tilde {q}w^{2}(\tilde{q}) \int _{y_{1}}^{y_{2}} \frac{\mathrm{d}y }{\sqrt{(2\tilde{k}\tilde{q})^{2} - (y-\tilde{E}-1+\tilde{k}^2+\tilde{q}^2)^{2}}}  
			\frac{\Im m \ \chi(\tilde{q},y)[\Theta(y)-\Theta(y-\tilde{E})] }{\left[1-w(\tilde{q}) \Re e \ \chi_{0}(\tilde{q},y)\right]^2 + \left[w(\tilde{q})\Im m \ \chi_{0}(\tilde{q},y)\right]^2}.
		\end{split}
		\ee
	\end{widetext}
	where $y_1 =\tilde{E} -(\tilde{k}+\tilde{q})^2+1 $, $y_2 =\tilde{E} -( \tilde{k}-\tilde{q})^2+1 $ and $ \tilde{\Omega}_{ZS} = \hbar\Omega_{ZS} / \varepsilon_{F}$.

\subsection{The behavior of $\Im m \, \Sigma (k = k_{\rm F} , E)$ in the vicinity of Fermi energy}\label{sec:app3}
\begin{figure*}
	\centering
\includegraphics[width=1.0\textwidth]{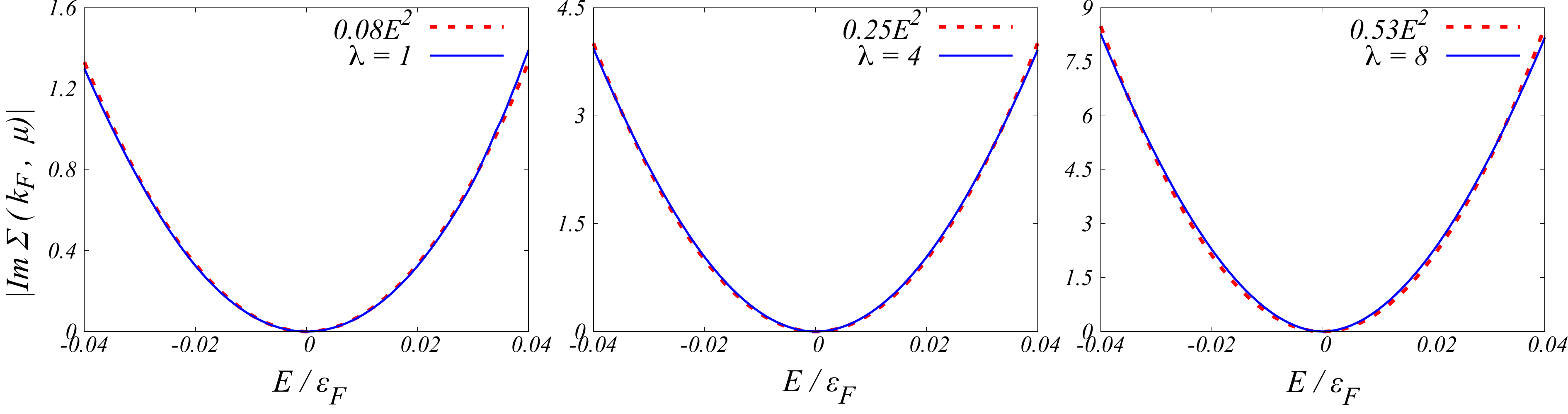} 
\caption{The imaginary part of self-energy (in units of $10^{-4} \varepsilon_{\rm F}$) calculated  for different values of the coupling constant at $k = k_{\rm F}$, and in the vicinity of the chemical potential. 
The red dashed lines are the parabolic fit to the imaginary part of self-energy for $E \rightarrow \mu $.
\label{fig:ImE}}
\end{figure*}

The numerical solution of the imaginary part of self-energy for $k=k_{\rm F}$ and $E \to \mu$ has been illustrated in Fig.\,\ref {fig:ImE}.
This figure shows that the quasiparticle decay rate vanishes as  $\propto (E -\mu)^{\beta}$ for $E \rightarrow \mu $ at $k \rightarrow k_{F}$. We have fitted our data into a parabolic function. It appears that at weak and intermediate couplings the $\beta= 2$ power provide an adequate fit to the numerical data. At strong couplings, a slightly smaller $\beta$ would provide even a better fit but still, we would find $1 < \beta \leq 2$. 

\begin{thebibliography}{90}
%
\bibitem{Baranov1}
M. A. Baranov, Phys. Rep. {\bf 464}, 71 (2008).
%
\bibitem{Baranov4}
M. A. Baranov, M. Dalmonte, G. Pupillo, and P. Zoller, Chem. Rev. {\bf 112}, 5012 (2012).
%
\bibitem{Bloch}
I. Bloch, Nature Phys. {\bf 1}, 23 (2005).
%
\bibitem{Matveeva}
N. Matveeva and S. Giorgini, Phys. Rev. Lett. {\bf 109}, 200401 (2012).
%
\bibitem{Parish}
M. M. Parish and F. M. Marchetti, Phys. Rev. Lett. {\bf 108}, 145304 (2012).
%
\bibitem{Abedinpour}
S. H. Abedinpour, R. Asgari, B. Tanatar, and M. Polini, Ann. Phys. (N.Y) {\bf 340}, 25 (2014).
%
\bibitem{Yamaguchi}
Y. Yamaguchi, T. Sogo, T. Ito, and T. Miyakawa, Phys. Rev. A {\bf 82}, 013643 (2010).
%
\bibitem{Zinner}
N. T. Zinner and G. M. Bruun, Eur. Phys. J. D {\bf 65}, 133 (2011).
%
\bibitem{Sieberer}
L. M. Sieberer and M. A. Baranov, Phys. Rev. A {\bf 84}, 063633 (2011).
%
\bibitem{Lu1}
Z.-K. Lu and G. V. Shlyapnikov, Phys. Rev. A {\bf 85}, 023614 (2012).
%
\bibitem{Lu2013}
 Z.-K. Lu, S. I. Matveenko, and G. V. Shlyapnikov, Phys. Rev. A {\bf 88}, 033625 (2013).
%
\bibitem{Jalabert}
R. Jalabert and S. Das Sarma, Phys. Rev. B {\bf 40}, 9723 (1989).
%
\bibitem{Polini}
M. Polini, R. Asgari, G. Borghi, Y. Barlas, T. Pereg-Barnea, and A. H. MacDonald, Phys. Rev. B {\bf 77} 081411 (2008).
%
\bibitem{Hwang}
E. H. Hwang and S. Das Sarma, Phys. Rev. B {\bf 77}, 081412(R) (2008).
%
\bibitem{Bostwick}
A. Bostwick, F. Speck, T. Seylle, K. Horn, M. Polini, R. Asgari, A. H. MacDonald and E. Rotenberg, Science {\bf 328}, 999 (2010).
%
\bibitem{Godfrin}
H. Godfrin, M. Mescheko, H. J. Lauter, A. Sultan, H. M. B\"{o}hm, and E. Krotscheck, Nature {\bf 483}, 576 (2012); S. T. Bramwell and B. Keimer,
Nature Mat., {\bf 13}, 763 (2014).
%
\bibitem{mazzanti_prl2009}
F. Mazzanti, R. E. Zillich, G. E. Astrakharchik, and J. Boronat, Phys. Rev. Lett. \textbf{102}, 110405 (2009).
%
\bibitem{hufnagl_prl2011}
D. Hufnagl, R. Kaltseis, V. Apaja, and R. E. Zillich, Phys. Rev. Lett. \textbf{107}, 065303 (2011).
%
\bibitem{macia_prl2012}
A. Macia, D. Hufnagl, F. Mazzanti, J. Boronat, and R. E. Zillich, Phys. Rev. Lett. \textbf{109}, 235307 (2012).
%
\bibitem{Asgari1}
R. Asgari, B. Davoudi, M. Polini, G. F. Giuliani, M. P. Tosi, and G. Vignale, Phys. Rev. B {\bf 71}, 045323 (2005).
%
\bibitem{torma_scripta2016}
P. T{\"o}rm{\"a}, Phys. Scr. \textbf{91}, 043006  (2016).
%
\bibitem{Gupta2003}
 S. Gupta, Z. Hadzibabic, M. W. Zwierlein, C. A. Stan, K. Dieckmann, C. H. Schunck, E. G. M. van Kempen, B. J. Verhaar and W. Ketterle,  Science {\bf 300}, 1723 (2003).
%
\bibitem{Vignale}
G. F. Giuliani and G. Vignale, \textit{Quantum Theory of the Electron Liquid} (Cambridge University Press, Cambridge, 2005).
%
\bibitem{Asgari2}
R. Asgari and B.Tanatar, Phys. Rev. B {\bf 74}, 075301 (2006).
%
\bibitem{Qaiumzadeh}
A. Qaiumzadeh and R. Asgari, New. J. Phys. {\bf 11}, 095023 (2009).
%
\bibitem{Boronat}
J. Boronat, J. Casulleras, V. Grau, E. Krotscheck, and J. Springer, Phys. Rev. Lett {\bf 91}, 085302 (2008).
%
\bibitem{Seydi}
I. Seydi, S. H. Abedinpour, and B. Tanatar, J. Low Temp. Phys. {\bf 187}, 705 (2017).
%
\bibitem{Mahan}
G. D. Mahan, \textit{Many-Particle Physics} (Plenum Press, New York, 2000) 3rd ed.
%
\bibitem{Lundqvist}
B. I. Lundqvist, Phys. kondens. Materie {\bf 6}, 193 (1967).
%
\end{thebibliography}
\end{document}